\documentclass[aps,pra,preprint,groupedaddress,showpacs]{revtex4}

\usepackage{graphicx}
\usepackage{hyperref}
\usepackage[latin1]{inputenc}

\newcommand{\QPH}[1]%
    {\href{http://arxiv.org/abs/quant-ph/#1}{quant-ph/{#1}}}

\usepackage{layout}

\begin{document}


\title{Photoionisation loading of large Sr$^+$ ion clouds with ultrafast pulses.}

\author{S. Removille, R. Dubessy, Q. Glorieux, S. Guibal, T. Coudreau, L. Guidoni, and J.-P. Likforman}

\email{Jean-Pierre.likforman@univ-paris-diderot.fr}

 \affiliation{Laboratoire Matériaux et Phénomènes Quantiques,\\
Université Paris Diderot et CNRS, UMR 7162,\\ 
Bât. Condorcet, 75205 Paris Cedex 13, France}

\begin{abstract}
This paper reports on photoionisation loading based on ultrafast pulses of singly-ionised strontium ions in a linear Paul trap.
We take advantage of  an autoionising resonance of Sr neutral atoms to form Sr$^+$ by two-photon absorption of femtosecond pulses at a wavelength of 431~nm.
We compare this technique to electron-bombardment ionisation and observe several advantages of photoionisation.
It actually allows the loading of a pure Sr$^+$ ion cloud in a low radio-frequency voltage amplitude regime.
In these conditions up to $4\times10^4$ laser-cooled Sr$^+$ ions were  trapped.
 \end{abstract}

\maketitle%
\section{Introduction}

Samples of laser-cooled ions confined in electromagnetic traps play a prominent role in several domains related to atomic physics: quantum information\cite{reviews_qi}, metrology\cite{reviews_metr}, quantum optics\cite{quantum_opt}.
The traditional method used to load an ion trap with the desired species is to create the ions directly inside the trapping region by electron-bombardment (EB) of a neutral atomic beam.
As already underlined by several groups\cite{Gulde2001,Lucas2004,Brownnut2007}
this technique, while very flexible because it applies to any atomic (and molecular) species, has several serious drawbacks.
Firstly, the electron beam may charge some insulator present in the neighbouring of the trap affecting the trapping potential with a slowly varying non-controlled electric field.
This situation induces an excess of radio-frequency (RF) heating that imposes a frequent readjustment of compensation-voltages  \cite{Lucas2004,Brownnut2007}.
Secondly, the vacuum quality of the setup is deteriorated by the presence of a hot filament near the trap.
Finally, the very small cross-section of the electron-impact ionisation requires huge atomic fluxes that negatively affect both the vacuum and the electrode surfaces.
In recent years, several groups have developed photoionisation techniques that eliminate these drawbacks.
We can distinguish two different photoionisation methods, depending on the excitation path from the neutral atom to the ionised state.
In what we will call "two-step" photoionisation (TSPI) a first narrowband cw laser is tuned on an intermediate transition and the ionisation threshold is attained with a second photon (that can possibly  have the same energy).
This method needs a non-negligible population in
 the intermediate level (saturating stabilized laser).
An other technique consists in using  a two-photon transition that directly brings a neutral atom above the ionisation threshold (two-photon photoionisation, TPPI).
TPPI requires high peak-power, normally associated to short-pulse lasers with a large spectral width.

TSPI presents the important advantage of isotope-selectivity through the shifts associated to the intermediate transition. 
It has been applied to Mg \cite{Madsen2000,Kjaergaard2000}, Ca \cite{Kjaergaard2000,Gulde2001,Lucas2004}, Ba \cite{Steele2007}, Yb \cite{Ytterbium}, and Sr \cite{Brownnut2007}.
Cd$^+$ ions have been produced using short pulses \cite{Deslauriers2006}, taking advantage of  the pulse duration of Titanium:Sapphire (Ti:Sa) sources to efficiently double and quadruple the fundamental frequency.
These experimental results are interpreted in terms of TSPI, however when spectrally-large pulses excite long-lived intermediate state, TPPI might also occur.

This paper reports on photoionisation loading of Sr$^+$ in a linear Paul trap based on ultrafast pulses.
We demonstrate that TPPI presents indeed several advantages with respect to EB.
In particular it allowed us to selectively load the trap with Sr$^+$ ions, to explore trapping regimes with low RF voltages, to obtain larger ion clouds and to improve the vacuum quality by lowering the power in the Sr oven.
Such a lower atom flux is particularly interesting in the case of micro-fabricated traps because it prevents shortcuts on the electrodes and reduces heating associated to patch potentials\cite{DeVoe2002}.
Let us finally mention that, because of its two-photon character, TPPI has the potential advantage to define an extremely well-localized spatial region for the ion production associated with the waist of the ionising beam\cite{two_photon_micro}.
This facilitates the ion creation at the center of the trap with an initially low potential energy.      

\section{experimental setup}

The level structure of the Sr atom is similar to that of other atoms with two electrons on the external shell (the relevant levels are represented in Figure 1a).
The $^1P_1$ intermediate level has been used to implement TSPI technique in Sr\cite{Brownnut2007}, following the example of Ca\cite{Gulde2001} and Yb\cite{Ytterbium}.
In the particular case of the Sr atom, the presence of a $(4d^2+5p^2) ^1D_2$ autoionising state that lies above the ionisation threshold is very helpful to increase the photoionisation efficiency.
The spectral characteristics of this state have been investigated earlier by two-color spectroscopy\cite{Mende1995}, in particular a cross section of 5600 Mb for the $^1P_1\rightarrow (4d^2+5p^2) ^1D_2$ transition has been measured.
Such an autoionising state can also be reached from the ground state by two-photon absorption: the $(5s^2) ^0S_0\rightarrow (4d^2+5p^2) ^1D_2$ two-photon transition has a linewidth of 0.7 nm FWHM (due to the short lifetime of the final state) and can be resonantly driven by two photons centered at 431 nm \cite{Baig1998}.
The strategy of the TPPI loading experiment described in this paper is based on this transition (Figure 1b). 

The neutral Sr atomic beam is produced in an oven placed below the trap and formed by a tungsten filament in which flows a maximum current of 1.15~A, that corresponds to a heating power of 1.3~W.
The temperature of the (bare) filament display an approximate linear dependence on the current and we measured 110$^\circ$ C for 0.8~A and 170$^\circ$ C for 1.2~A.
In these conditions we expect an Sr partial pressure roughly in the range 10$^{-13}$ -- 10$^{-9}$ mbar\cite{pressure}.
An electron gun based on a thorium tungsten wire allows us to perform EB with a typical electron energy of $\simeq 300$~eV.
The photoionising laser source is based on a homemade femtosecond Ti:Sa oscillator 
that produces 50~fs pulses of 1.5~nJ energy at 862~nm with a repetition rate of 100~MHz.
The pulses are frequency doubled in a 0.5~$\mu$m thick Beta Barium Borate (BBO) nonlinear crystal.
The thickness of the crystal is such that group velocity mismatch negligibly affects pulse-duration.
A pulse energy of more than 0.15~nJ is routinely obtained at 431~nm.
The beam is directed into the vacuum chamber and focused at the center of the ion trap by a lens of focal distance 200~mm.
The measured photoionising beam size at the waist is $20\pm 2$~$\mu$m, corresponding to a peak intensity of $\simeq 1$~GW/cm$^2$. 

Sr$^+$ ions are trapped in a standard linear Paul trap (see Figure 2) \cite{Prestage1989}.
Four parallel rod electrodes (diameter 6.35~mm, inner radius $r_0 = 3.2$~ mm) are used for the radio-frequency (rf) radial confinement.
A 2.5~MHz rf potential with an amplitude $V_{rf}$  in the range 50--400~V is applied to two of the diagonally opposed electrodes.
The remaining two rods are normally grounded or used to resonantly excite the ion motion in the trap (see below).
The Sr$^+$ ion radial movement is defined by a typical secular frequency $\nu_R=400$~kHz for an applied voltage $V_{rf}=500$~V.
Two annular "end caps" separated by 20~mm are used for the longitudinal confinement and brought to a static positive voltage ($V_{ec}=500$~V).
The corresponding axial frequency is $\nu_A=20$~kHz.
Trapped ions can be detected in a destructive way by an ejection sequence: an asymmetric potential is applied to the "end caps" that kicks out the cloud along the trap axis through an electrostatic lens towards an electron multiplier.
This detection scheme allows us to perform ion-counting and to analyze the trapped species by performing mass-spectra.
A mass spectrum is obtained by measuring the losses induced in successive realizations of the cloud in the presence of a "tickle" sinusoidal excitation applied to two diagonally opposed rods.
The result of this parametric excitation is a depletion of the ion signal when the tickle frequency matches an integer fraction of $2\nu_R$ (negative peak in the mass spectrum)\cite{Guidoni2005}.

Trapped Sr$^+$ ions are Doppler cooled using the $^2S_{1/2}\rightarrow ^2P_{1/2}$ optical transition at 422~nm (natural linewidth, $\Gamma/2\pi= 20.2$~MHz, Figure 1c).
This transition is driven using laser light generated by a commercial extended-cavity diode laser (Toptica DL100).
A commercial "repumping" fiber-laser (Koheras Adjustik Y10) at 1092~nm drives the $^2D_{3/2}\rightarrow ^2P_{1/2}$ transition to avoid the accumulation of the ions (optical pumping) into the metastable $^2D_{3/2}$ state.
Let us mention that this setup allows us to perform both the measurement of the ion fluorescence, useful for optimizing the cooling phase, and the precise evaluation of the number of trapped ions by ejection and ion-counting detection.
An example of a fluorescence image of a large cloud (2~mm length and 200~$\mu$m width, FWHM) containing $3\times 10^4$ Sr$^+$ ions is shown in Figure 3.
In order to remove background light from images, we perform spatial filtering in an intermediate object plane through a 10~mm long and 0.5 mm wide slit and we filter both in polarization and wavelength (interferential filter) the detected photons.

\section{Results}
In a first experiment, we performed mass spectra of the trapped ions produced either by EB or by TPPI.
Typical mass spectra corresponding to these techniques are shown on Figure 4.
Basically two main peaks associated to Sr$^+$ are expected and observed, corresponding to a tickle frequency of $2\nu_R$ and $\nu_R$ (negative peaks in the mass spectrum).
This frequency (400~kHz for the experimental conditions of Figure 4) depends on the trap potentials and the ion mass.
Secondary satellite peaks of lower amplitude separated by $\nu_A\simeq 40$~kHz are visible at the second harmonic frequency $2\nu_R$ (800~kHz) that originate from the coupling between the axial and longitudinal motion in the large ion clouds\cite{Vedel1998, Dubessy2007}.
The main difference between the two spectra is the value of the peak contrast.
The contrast of the saturated peak at $2\nu_R$ gives us information about the proportion of Sr$^+$ ions in the trap. TPPI produces a contrast of 100\%, indicating that a pure Sr$^+$ cloud is obtained, compared to 66\% in the case of EB for which other ion species are also produced.
As we demonstrated that only strontium ions are produced in the trap when using TPPI, we also confirmed that this ionisation process is actually based on two-photon absorption.
Indeed, the ionisation rate as a function of the ionising beam intensity shows the expected quadratic dependence (Figure 5).
The reported rates are obtained by performing several load-eject cycles for different short loading times $\tau$ (see below), and by extracting the slope of the linear behavior by a fit.
The error bars represent the average variance as extracted by the fit procedure.

In a second experiment, we compared the loading times and the maximum number of trapped ions in the two cases of EB and TPPI.
The experiment is performed as follows: we turn-on the electron gun or the photoionising beam at time $t=0$ ; after a certain time $\tau$, we turn off the laser cooling and eject the loaded ions towards the ion counter.
Laser cooling is stopped in the ejection phase to avoid spurious losses induced by damping.
We show in Figure 6a the number of trapped ions, as function of $\tau$, obtained using the two techniques. 
In both cases, saturation in the number of trapped ions is reached within a few tens of seconds.
However, only TPPI allowed us to create pure $Sr^+$ clouds containing a large number of ions: the EB technique saturates around 1000 trapped ions.
In fact, with TPPI an optimum of $4\times 10^4$ trapped Sr$^+$ ions is obtained for $V_{rf}=125$~V, that is for a relatively shallow trap.
In the case of EB it is impossible to explore the same range for the parameter $V_{rf}$ without affecting the sample purity, since too many spurious ionic species are produced and trapped in shallow potential wells.
This effect is clearly visible in the mass spectra of Figure 6b obtained with the two techniques in an intermediate regime ($V_{rf} = 350$~V).
Therefore, in our case of a non-crystalized regime, the possibility to work at a low $V_{rf}$  is crucial in order to obtain a large number of trapped ions.
We interpret this result in terms of rf-heating process that increases with $V_{rf}$\cite{rf_heating}.
In this situation, and in the presence of laser-cooling, the ion-density depends on the balance between cooling and heating.
A maximum density is then achievable by minimizing $V_{rf}$ (i.e. the rf heating) within the stability region.
But the total number of trapped ions also depends on the volume of the trap that is imposed by $V_{rf}$ and $V_{ec}$. 
Numerical simulations carried-out using Simion\textregistered~  software\cite{simion} allowed us to estimate the trap volume as a function of $V_{rf}$ for $V_{ec}=500$~V.
An optimum (maximum) volume is obtained for $V_{rf}=180$~V, slightly larger than the experimental value that maximizes the total number of trapped ions.

We also observed that, when working in the regime of relatively high $V_{rf}$, that is optimal for EB ($V_{rf}= 500V$), 1100 ions can be trapped by EB to be compared to 1200 ions obtained by photoionisation.
This small discrepancy might be explained by space charge effects due to electrons that can perturb the trap electrostatic potential or by the intrinsic possibility, and advantage, of photoionisation to produce ions at the centre of the trap.
Let us also finally remark that the selectivity of the TPPI technique allowed us to load pure $Sr^+$ clouds in the trap at very low oven temperatures (on the order of $120^\circ$~C), contrary to the case of EB. 

\section{Conclusion}

We demonstrated two-photon photoionisation loading of Sr$^+$ in a linear Paul trap using ultrafast pulses centered at 431~nm.
We compared this technique to the electron bombardment loading and observed several advantages, already mentioned in previous experiments concerning other species or other photo-excitation paths.  
In particular this technique allowed us to selectively load pure $Sr^+$ clouds, to explore trapping regimes with low RF voltages, to obtain large cooled-ion clouds, and to improve the vacuum quality by lowering the power in the Sr oven.
Our results have been obtained with a home-made femtosecond source that delivers relatively low energy pulses.
We expect an improvement of two orders of magnitudes in the photoionization cross-section in the case of commercially available Ti:Sa oscillators delivering routinely 10 nJ per pulse.  
Let us finally mention that the trapping of up to $4\times10^4$ cooled Sr$^+$ is an important step towards the realization of an ion-based quantum memory\cite{nous} 

\section{Acknowledgments}
We thank P.~Lepert for technical support
.
The authors would also like to thank M. Joffre for the lending of the femtosecond oscillator.
This work was supported by ANR "jeunes chercheuses et jeunes chercheurs" research contract JC05\_61454.

\newpage
\begin{figure}[h]
\centerline{\scalebox{0.6}{\includegraphics{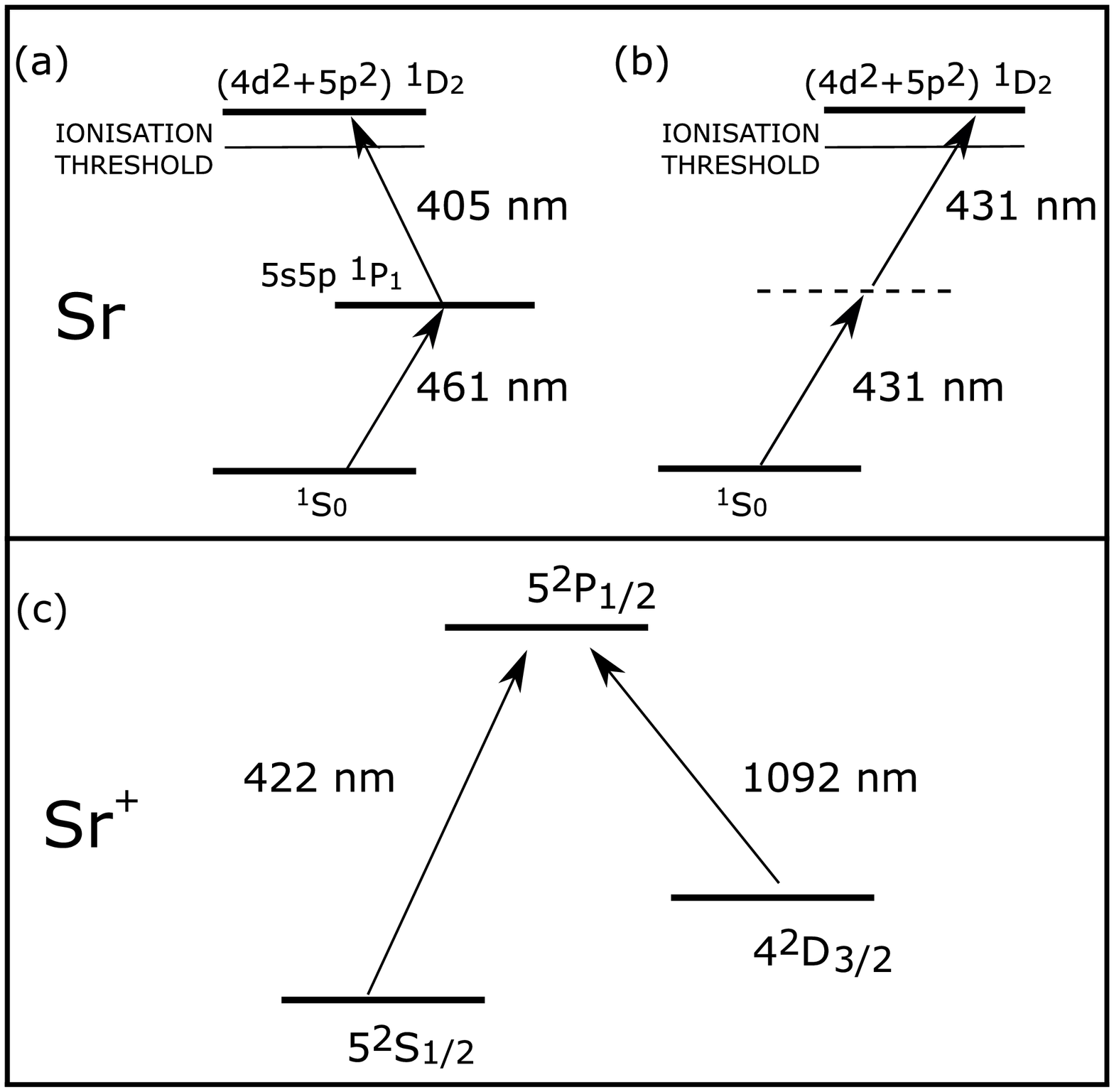}}}
  \caption{Relevant energy levels of neutral Strontium involved in: a) two-step photoionisation TSPI;  b) two-photon photoionisation TPPI discussed in this paper. c) Relevant energy levels of $Sr^+$ ion involved in Doppler cooling.}
\end{figure}
\begin{figure}[h]
\centerline{\scalebox{0.6}{\includegraphics{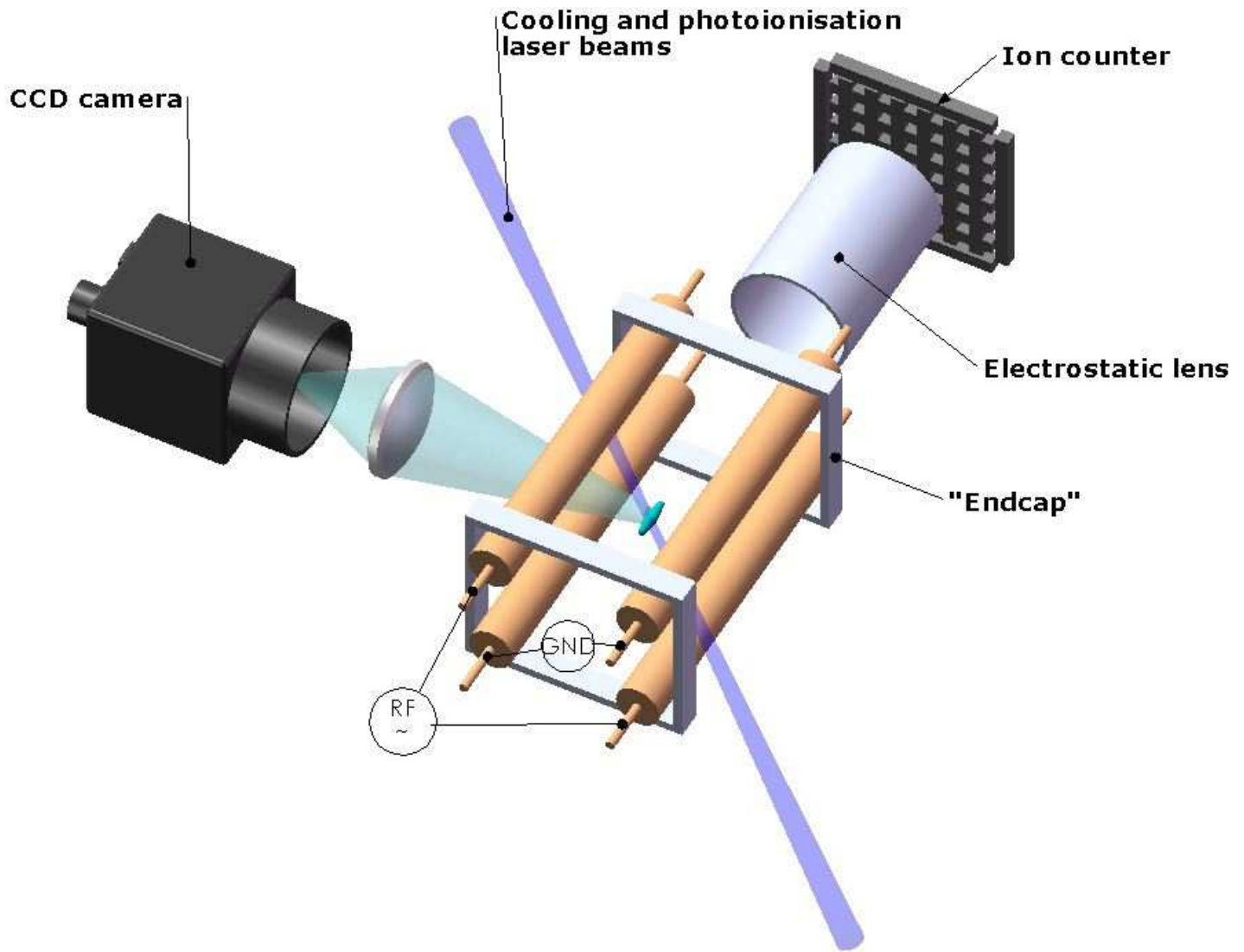}}}
 \caption{Schematic description of the linear Paul trap setup.  
Four parallel rod electrodes are used for the rf radial confinement. Two annular "end caps" separated by 20~mm are used for the longitudinal confinement. The trapped ions can be axially ejected and counted by an electron multiplier. The ion cloud fluorescence resulting from Doppler cooling is recorded on a ccd camera.}
\end{figure}
\begin{figure}[h]
\centerline{\scalebox{0.6}{\includegraphics{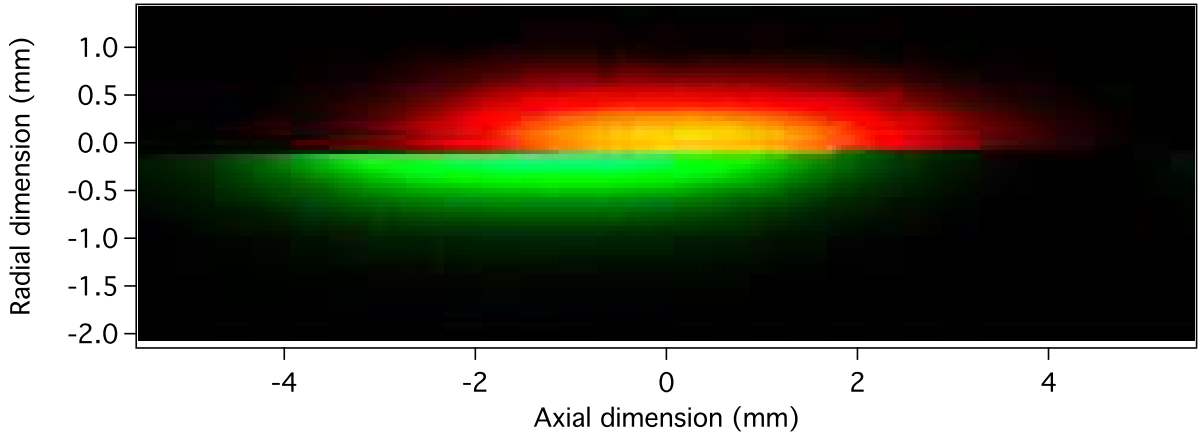}}}
  \caption{Fluorescence image at 422~nm of a trapped ion cloud (2~mm length and 200~$\mu$m width, FWHM) containing $3\times 10^4$ Sr$^+$ ions ($V_{rf}=130$~ V, $V_{ec}=500$~V).}
\end{figure}
\begin{figure}[h]
\centerline{\scalebox{0.5}{\includegraphics{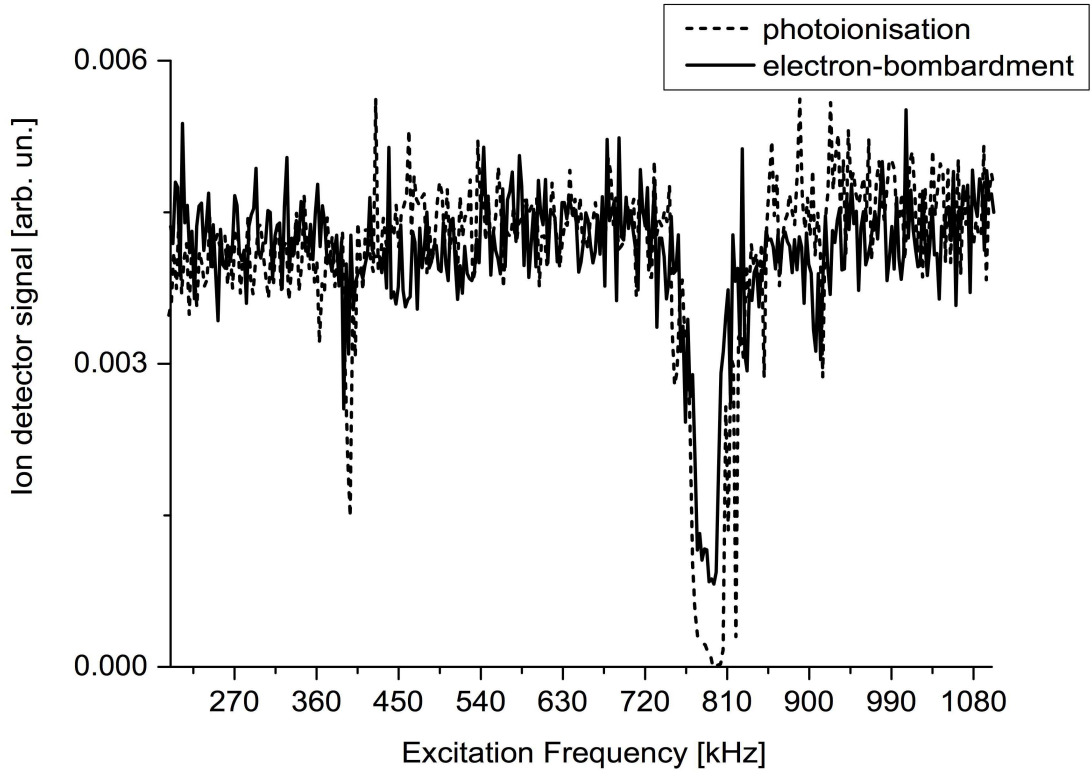}}}
  \caption{Comparison of mass spectra of the trapped ions produced either by electron-beam or by photoionisation.
  The two main peaks associated to Sr$^+$ are observed for a tickle frequency of $2\nu_R$ and $\nu_R$ (negative peaks in the mass spectrum). $V_{rf}=500$~V.}
\end{figure}
\begin{figure}[h]
\centerline{\scalebox{0.4}{\includegraphics{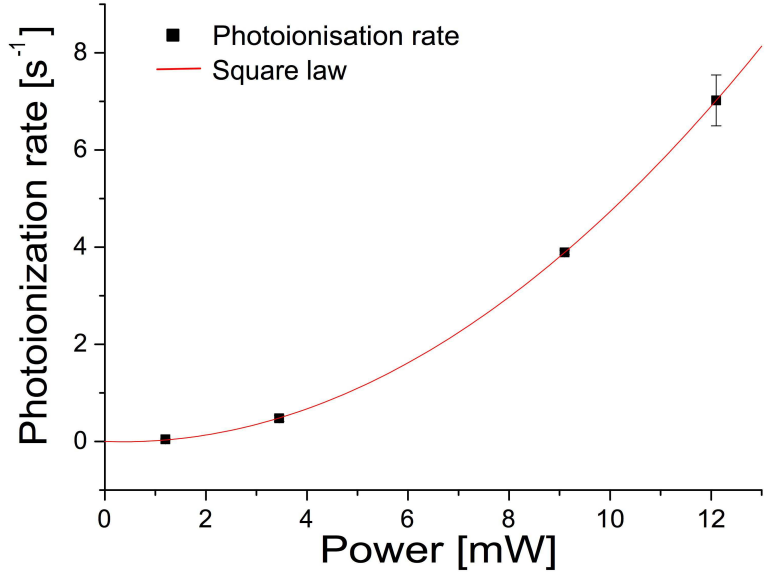}}}
  \caption{Photoionisation rate as function of the average power of the ionisation beam obtained for $V_{rf}=130$~V and $V_{ec}=500$~V. An excellent quadratic law is observed, as expected for a two-photon absorption process.}
\end{figure}
\begin{figure}[h]
\centerline{\scalebox{0.3}{\hfil\includegraphics{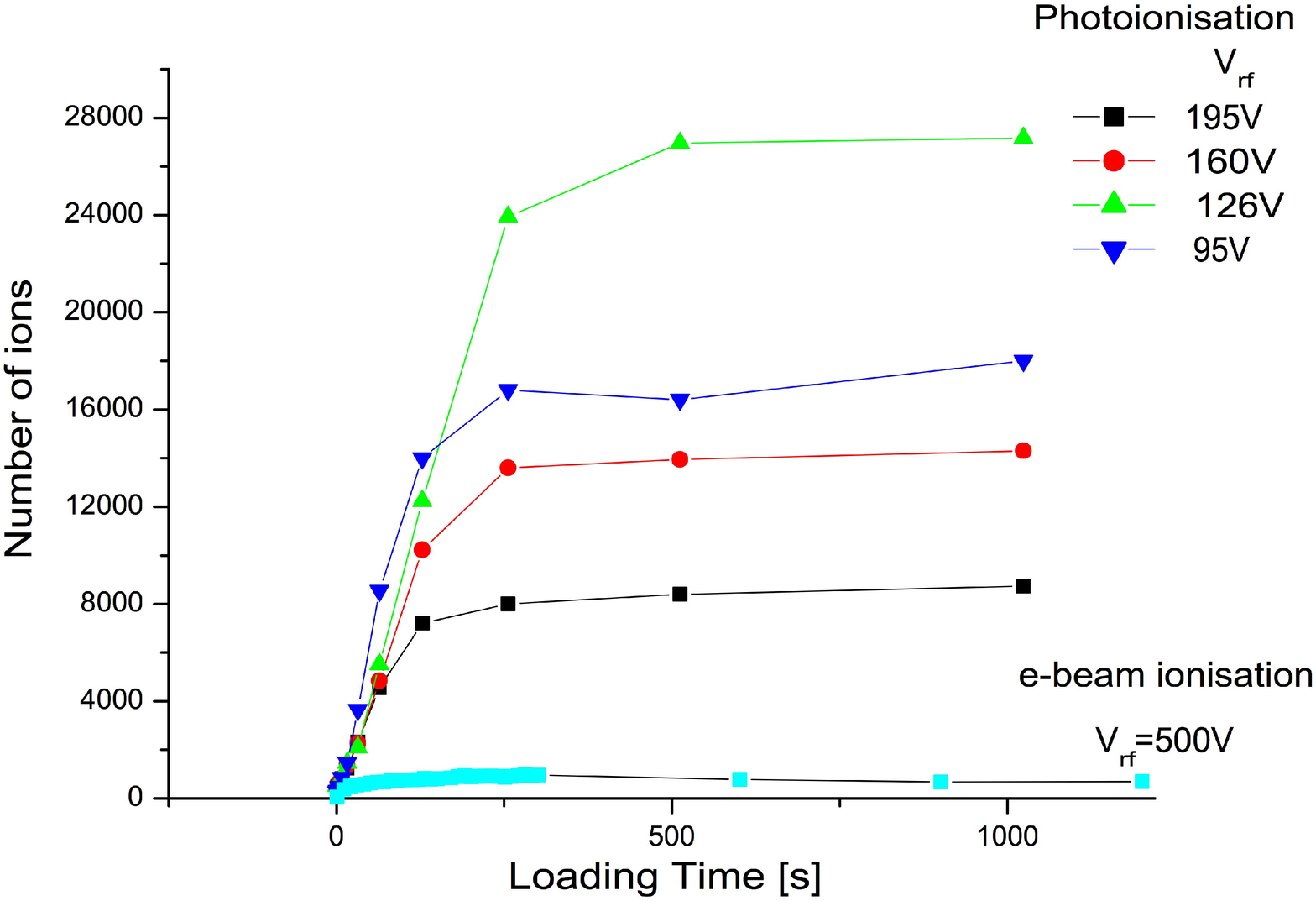}\hfill\includegraphics{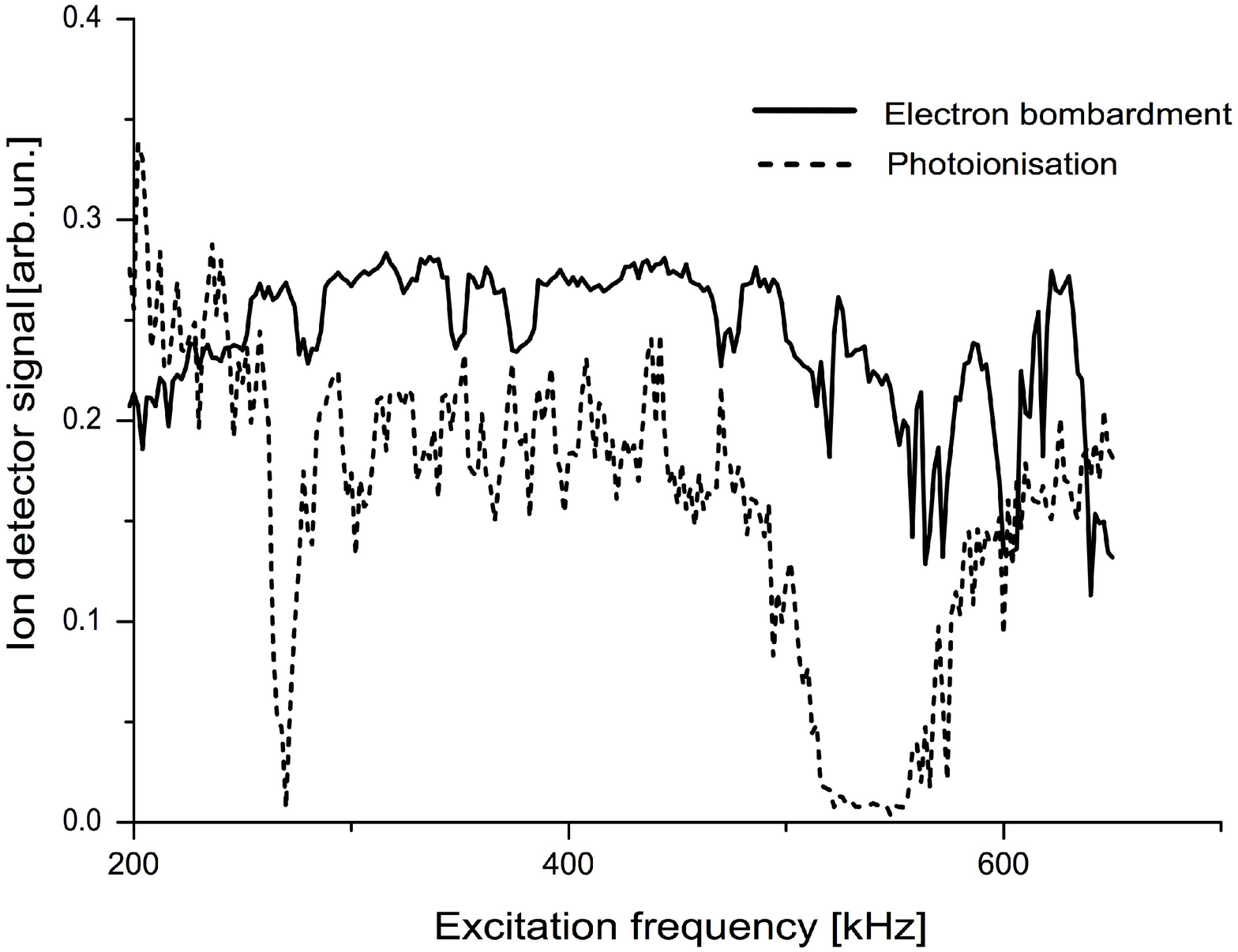}\hfil}}
  \caption{a) Number of trapped ions as function of the loading time. In the case of photoionisation, the curves corresponding to several $V_{rf}$ are traced. b) Mass spectra of the trapped ions produced either by electron-beam (solid curve) or by photoionisation  (dashed curve) for $V_{rf}=350$~V. All the data are obtained for $V_{ec}=500$~V. 
}
\end{figure}

\end{document}